\documentstyle[preprint,prb,aps]{revtex}
\tighten
\begin{document}
\title{Detailed electronic structure studies on superconducting MgB$_2$ and related compounds}
\author{P.Ravindran, P.Vajeeston, R.Vidya, A.Kjekshus and H.Fjellv{\aa}g}
\address{$^1$ Department of Chemistry, University of Oslo, Box 1033, Blindern,
N-0315, Oslo, Norway\\
}
\date{\today}
\maketitle
\begin{abstract}
{Our recent electronic structure studies on series
of transition metal diborides indicated that the electron phonon coupling constant is much smaller in
these materials than in superconducting intermetallics. 
However experimental studies recently show an exceptionally large superconducting transition
temperature of 40\,K in MgB$_2$. In order to understand the unexpected superconducting behavior
of this compound we have made electronic structure calculations for MgB$_2$ and 
closely related systems.
Our calculated Debye temperature from the elastic properties indicate that the average
phonon frequency is very large in MgB$_2$ compared with other superconducting 
intermetallics and the exceptionally high $T_c$ in this material can be explained through
BCS mechanism only
if phonon softening occurs or the phonon modes are highly anisotropic.
We identified a doubly-degenerate quasi-two
dimensional key-energy 
band in the vicinity of $E_{F}$ along $\Gamma$-A direction of BZ (having equal amount of 
B $p_{x}$ and $p_{y}$ character) which play an important role in deciding the 
superconducting behavior of this material. Based on this result, we have searched 
for similar kinds of electronic
feature in a series of isoelectronic compounds such as BeB$_2$, CaB$_2$, SrB$_2$, LiBC and
MgB$_2$C$_2$ and found that MgB$_2$C$_2$ is one potential 
material from the superconductivity point of view. We have also investigated closely related compound
MgB$_4$ and found that its $E_{F}$ is lying in a pseudogap with a negligibly small density
of states at $E_{F}$ which is not favorable for superconductivity. There are contradictory
experimental results regarding the anisotropy in the elastic properties of 
MgB$_2$ ranging from
isotropic, moderately anisotropic to highly anisotropic. In order to settle this issue
we have calculated the single crystal elastic constants for MgB$_2$ by the accurate 
full-potential method and derived the directional dependent linear compressibility,
Young's modulus, shear modulus and relevant elastic properties from these results. 
We have observed large anisotropy in the elastic properties 
consistent with recent high-pressure measurements.
Our calculated polarized optical dielectric tensor shows highly anisotropic behavior 
even though it possesses isotropic transport property. MgB$_2$ possesses a mixed bonding
character and this has been verified from density of states, charge density
and crystal orbital Hamiltonian population analyses.
}
\end{abstract}
\section{INTRODUCTION}
\label{sec:intro}
The recent discovery of superconductivity with high $T_{c}$\cite{nagamatsu01} in MgB$_2$
has initiated large activity in experimental as well as theoretical studies. 
Nb$_3$Ge is long been the record holder of the highest $T_{c}$ among the intermetallic 
superconductors. The recently found YPd$_2$B$_2$C touches the same $T_{c}$ as 
Nb$_3$Ge. 
The work of Bednorz and M\"{u}ller\cite{bednorz86} in 1986 on the basis of copper oxides
started the discovery of a rapidly increasing number of high-$T_{c}$ superconductors, among 
which the current
record is held by HgBa$_2$Ca$_2$Cu$_3$O$_{8-\delta}$ with $T_{c}$$\sim$ 164\,K 
under pressure.\cite{gao94}
During the past decade, remarkable progress in basic research and technological
applications has been made on the high-$T_{c}$ cuprate superconductors. However the complex
crystal structures in combination with the multicomponent nature of the materials 
involved hinder the full understanding
of microscopic origin of high-$T_{c}$ superconductivity. Hence it may be beneficial to 
study the properties
of a simple compound like MgB$_2$ in detail, which does not only have high $T_{c}$, but also takes 
a simple crystal structure with $sp$ electrons (involved in the  
superconducting process) that are easy to handle theoretically.   
\par
In cuprates it is now generally believed that the conductivity takes place in the CuO$_2$ 
planes which,
therefore, are essential to the high-$T_{c}$ superconductors. The observation of 
superconductivity with high $T_{c}$  
in MgB$_2$ releases the question of whether superconductivity with even higher
$T_{c}$ can be found in intermetallics which do not comprise the characteristic CuO$_2$ planes.
Several properties of MgB$_2$ appear closely related to high $T_{c}$ superconducting cuprates:
a low electron density of states, a layered 
structural character and the presence of rather
light atoms (like oxygen in cuprates) facilitate high phonon frequencies. Further, although 
the electronic structure of
this material has three-dimensional character, the B-B $\sigma$ bands derived from B $p_{x,y}$ electrons
(believed to be important for the superconductivity) reflect two-dimensional character.
Like the cuprates, MgB$_2$ appears to exhibit hole conductivity, as evidenced from theoretical
considerations\cite{kortus01,hirsch01,belashchenko01,an01} and experimental Hall coefficient
measurement.\cite{hall}
Another interesting aspect of the MgB$_2$ structure is that it has negatively charged 
honeycomb-shaped B planes
which are reminicent of highly negatively charged Cu-O planes in the high-$T_{c}$ cuprates.
\par
It is often believed that the $d$ electrons play an important role for the
superconducting behaviour of intermetallic compounds. So the experimental search for new
superconductors has to a large extent been focussed on transition metal compounds. 
The back-ground for this is that transition metal compounds
usually possess large density of states (DOS) at the Fermi level (one of the ingredients 
required for high $T_{c}$) than the main-group metal ($sp$) compounds.
The BCS superconductors usually possess larger
DOS at the Fermi level ($E_{F}$) and the high $T_{c}$ in conventional superconductors is 
related to the large  
$N(E_{F})$ values as well as the strong coupling of selected phonon modes to the electronic system. 
Further, relatively high T$_c$ in superconducting intermetallics are believed to 
be coupled to a van Hove-like peak 
at $E_{F}$ in the DOS curve, as predicted by various bandstructure calculations and indicated experimentally
for Ni-site substituted rare-earth nickel borocarbides.\cite{gangopadhyay95} 
However in MgB$_2$ there is no such
peak-like feature at the vicinity of $E_{F}$ in the DOS profile and hence one must search for some other
origin of the superconductivity.  MgB$_2$ is
a $sp$ metal and hence its value of $N(E_{F})$ is small compared with that of superconducting
transition metal compounds. 
Further, the presence of light borons in MgB$_2$ similar to oxygens in high-$T_{c}$ cuprates,   
indicates that some exotic mechanism is involved in the
superconductivity. On the other hand, theoretical studies\cite{an01} show that the 
quasi two-dimensional B $\sigma$ bands are strongly coupled with the $E_{2g}$ phonon modes 
which would be consistent with the BCS mechanism. 
The experimentally observed isotope effect\cite{budko01} in MgB$_2$ also indicates
phonon-mediated
superconductivity. But, as discussed by Baskaran,\cite{baskaran01} the absence of Hebel-Slichter 
peak in the NMR relaxation,\cite{kotegawa01,gerashenko01,jung01}
a temperature dependent peak around 17\,meV in the energy resolved neutron scattering,\cite{sato01}
first order metal-to-metal transition on Al or C substitution,\cite{slusky01,ahn01,takenobu01}
apparently anomalous temperature dependences of the Hall coefficient\cite{kang01} and London
penetration depth\cite{panagopoulos01} and the need of small $\mu^{*}$ value to explain
the experimentally observed high $T_{c}$ by the BCS theory\cite{kong01} indicate that the
mechanism differs significantly from that of the BCS theory.  
\par
The discovery of new classes of quaternary intermetallic superconductors; the 
borocarbide\cite{nagarajan94,cava94} and the 
boronitride\cite{cava941}
series, with relatively high $T_{c}$ (up to 23 K for YPd$_2$B$_2$C\cite{cava94})
has encouraged the search for superconductivity in materials possessing light atoms such
as B, C, N, H etc.  
Superconductivity  with the T$_c$
ranging 2$-$4 K has been observed for ternary transition metal diborides 
(YRe$_2$B$_2$,LuB$_2$C$_2$, YB$_2$C$_2$)\cite{chinchure00,philips89} 
and it is also observed in magnetic rare-earth (RE) rhodium borides 
(RERh$_4$B$_4$\cite{sinha89}). 
Very recently, the layer-structure compounds $\beta$-ZrNCl and $\beta$-HfNCl have been 
found to be superconducting upon Li intercalation with $T_{c}$ = 12.5 
and 25.5 K, respectively.\cite{yamanaka98}
Another group of materials (maximum $T_{c}$ = 9.97\,K reported\cite{kremer99} for Y$_2$C$_2$I$_2$)
is RE carbide halide superconductors with C$-$C pairs located in
octahedrally coordinated voids of close-packed RE atoms. 
The recent discovery of superconductivity\cite{sanfilippo00} at $\approx$14\,K
in the AlB$_2$-type phase of CaSi$_2$ (the highest $T_{c}$ ever obtained among the silicides)
indicates that the AlB$_2$-type structure may be favorable for superconductivity.
\par
There are several mechanisms proposed for high $T_{c}$ superconductivity in MgB$_2$ and the
like. One is based on band
structure findings,\cite{kortus01,kong01} which suggests that the superconducting state
results from strong electron-phonon
interaction and high phonon frequency associated with the light boron atom. This
is supported by the recent observation of relatively
large boron isotope effect on $T_{c}$.\cite{budko01} 
A recent high-pressure study\cite{tomita01} shows that $T_{c}$ decreases with pressure at a rate
of $-$1.11\,K/GPa which is consistent with the BCS framework.
It has been widely speculated that the low mass of boron is conducive for the occurrence
of high phonon frequencies and consequently for high $T_{c}$.
Another mechanism is called the "universal" by Hirsch\cite{hirsch01}, and this conjectures
that the superconductivity in MgB$_2$ (similar to that in cuprate superconductors) is driven by
the pairing of the heavily dressed holes in bands that are almost full to gain enough kinetic
energy to overcome the Coulomb repulsion. A positive pressure effect on $T_{c}$ has also
been predicted by Hirsch when the pressure reduces the interatomic B$-$B distance. 
An and Pickett\cite{an01} maintain that the B $\sigma$ bands are playing an important role
in the superconductivity of MgB$_2$, and that the B in-plane $E_{2g}$ phonon mode is strongly
coupled to this band. Contradictory to the above viewpoint, Baskaran\cite{baskaran01}
concluded from the resonance valence bond (RVB) theory that the two dimensional $\sigma$ bands do
not play a crucial role in establishing high $T_{c}$ superconductivity and
$p\pi$ band might interfere with superconductivity.
\par
MgB$_2$ possesses a hexagonal crystal structure and hence one can expect anisotropy in the physical
properties of this material.
Despite the strongly anisotropic layered hexagonal structure,
its electronic properties indicate three dimensionality and the calculated Fermi velocity
in the plane and perpendicular to the plane are almost the same.
The theory for hole superconductivity\cite{hirsch01} suggests that a decrease in the
B-B interatomic distance should increase $T_{c}$. The RVB theory\cite{baskaran01} suggests that
a increase in chemical pressure along the $c$ axis should decrease $T_{c}$ and ultimately
convert the material to a normal metallic state.  In order to test these possiblities it is 
important to know the anisotropy in mechanical properties.
From high-pressure total energy studies Loa and Syassen\cite{loa01} concluded
that MgB$_2$ has isotropic compressibility. From a high-resolution x-ray powder
diffraction high-pressure study along with the density functional calculation Vogt
{\em et al.}\cite{vogt01} concluded that MgB$_2$ possesses nearly isotropic mechanical
behavior. From isothermal compressibility measurements by 
synchrotron x-ray diffraction Prassides\cite{prassides01} concluded that MgB$_2$ is 
a stiff tightly-packed
incompressible solid with only moderate bonding anisotropy between inter- and intralayer
directions.
However, Jorgensen {\em et al.}\cite{jorgensen01} found unusually large anisotropy
in thermal expansion and compressibility from the neutron diffraction measurements.
So, it is interesting to calculate the single-crystal elastic constants of MgB$_2$
to identify the exact nature of the anisotropy.
\par
Several attempts have been made to enhance $T_{c}$ by substitution of such as
Al,\cite{slusky01} Be,\cite{felner01}, Zn\cite{kazakov01} and Li,\cite{zhao01} for Mg and C\cite{ahn01} for B but no
practical progress has hitherto been obtained. 
Moreover, the role of Mg and B site substitution on the electronic structure of MgB$_2$ has been
studied theoretically.\cite{medvedeva01}
So, it is interesting to search for compounds with an
electronic structure similar to that of MgB$_2$. A systematic investigation of this and related
materials is a way to learn the mechanism of superconductivity in this novel material and to
identify materials with high T$_c$.
Knowledge of electronic structure, DOS, Debye temperature and related properties
are important for assessing the mechanism and nature of superconductivity.
In a search for superconducting diborides, BeB$_2$ is a promising candidate
since the lighter Be may help to provide larger phonon frequencies and hence
increase $T_{c}$. If the electron per atom ratio is important for the superconductivity
in this class one should also consider CaB$_2$ and SrB$_2$ as interesting candidates. 
If it is the combination of Mg and B which brings the superconductivity in MgB$_2$,  one 
has to consider MgB$_4$ also. If the number of electrons in the B layers is a key for 
superconductivity in MgB$_2$, related layer-structured materials such as LiBC, MgB$_2$C$_2$ 
should be paid attention. For these reasons we have made detailed electronic structure studies
for the above mentioned compounds.
\par
The rest of the paper is organized as follows. The structural aspects and the computational
details about the calculations of the electronic structure, optical spectra and the
elastic constants are given in in Sec.\,\ref{sec:details}.
In Sec.\,\ref{sec:resdis} we have analyzed the bonding behavior of MgB$_2$ using
the orbital and site projected DOS, crystal orbital overlap Hamiltonian population 
(COHP), charge density analysis etc. The electronic band structure of MgB$_2$ is
calculated and compared with that of closely related systems and also analyzed the
possible connection between the electronic structure and the superconductivity.
The elastic and the optical anisotropy of this material are calculated and 
compared with available experimental results.  Finally we 
summarize the important findings of the present study in Sec.\,\ref{sec:con}.

\section{Structural aspects and computational details}
\label{sec:details}
\subsection{Crystal structure details}
\label{ssec:str}
MgB$_2$ (Fig.\,\ref{fig:str}a) 
has AlB$_2$-type structure\cite{jones54} 
with space group {\em P6/mmm} and lattice parameters, $a$ = 3.084\,\AA 
and $c$ = 3.522\,\AA. It is a simple hexagonal lattice 
of close-packed Mg layers alternating with graphite-
like B layers, viz. B atoms arranged at the corners of a hexagon 
with three nearest neighbor B atoms in each plane. The Mg atoms
are located at the center of the B hexagon, midway between
adjacent B layers. 
\par
MgB$_2$C$_2$ crystallizes in an orthorhombic structure,\cite{nesper1}
space group $Cmca$ with $a$ = 10.92, $b$ = 9.46 and $c$ = 7.45\,\AA.
The structure of MgB$_2$C$_2$ (Fig.\,\ref{fig:str}b) contains
graphite-like but slightly puckered boron-carbon layers whose charge
is counter-balanced by Mg$^{2+}$ cations. The mutual coordination of boron
and carbon consists of five atoms of the other kind, three of which being located in the 
same and two in adjacent layers. Each of Mg is
coordinated by six B and six C atoms arranged at the corners of a slightly distorted
hexagonal prism. 
The B-C distances within the layers range from 1.562 to 1.595\,\AA.
\par
LiBC (Fig.\,\ref{fig:str}c) crystallizes\cite{nesper2} in a hexagonal primitive lattice 
with space group $P$6$_3$/$mmc$.
The lattice parameters are $a$ = 2.752 and $c$ = 7.058\,\AA.
The B and C atoms form a planar so-called heterographite layer. The interlayer regions are
filled by Li. The B-C distance of 1.589\,\AA in LiBC is comparable with that in
MgB$_2$C$_2$.
\par
In MgB$_4$, the B atoms form interconnected pentagonal pyramids with the Mg atoms
located in channels running parallel to the $c$ axis. The Mg atoms form zig-zag chains.
MgB$_4$ is orthorhombic, space group $Pnma$, with $a$ = 5.46, $b$ = 7.47 and
$c$ = 4.42\,\AA.\cite{vegas95}. The average B-B distance in the pentagonal pyramid is
1.787\,\AA. Whenever possible we have used
the experimental lattice paramenters for our calculation, whereas for BeB$_2$,
CaB$_2$ and SrB$_2$ we have used the optimized structural parameters obtained
from total energy minimization.

\subsection{Computation details for the FPLAPW calculations}
These investigations are based on {\em ab initio} electronic structure calculations
derived from density-functional theory. For the 
screened plasma frequency and the orbital projected DOS calculations we have
applied the full-potential linearized-augmented plane wave (FPLAPW) method\cite{wien} 
in a scalar-relativistic version without spin-orbit (SO) coupling. In the calculation we 
have used atomic sphere radii 1.8 and 1.5 a.u. for Mg
and B, respectively.  
The charge density and
the potentials are expanded into lattice harmonics up to $\ell$ = 6 inside the spheres and
into a Fourier series in the interstitial region. 
The initial basis set included 3$s$, 3$p$ and 3$d$ valence and 2$s$, 2$p$ semicore
functions at the Mg site, 2$s$, 2$p$ and 3$d$ valence functions for the B site.
The set of basis functions was supplemented
with local orbitals for additional flexibility in representing  
the semicore states and for relaxing the linearization errors generally. 
The effects of exchange and correlation are treated within the
generalized-gradient-corrected local-density approximation using the
parameterization scheme of Perdew {\em et al.}\cite{pw96} To ensure 
convergence for the Brillouin
zone (BZ) integration, 320\,{\bf k}-points in the irreducible wedge of the first 
BZ of the hexagonal lattice for MgB$_2$ were used.
Self-consistency was achieved by demanding the convergence of the total
energy to be smaller than 10$^{-5}$\,Ry/cell. This corresponds to a convergence of the
charge below 10$^{-4}$ electrons/atom. For
BeB$_2$, CaB$_2$ and SrB$_2$ we have made the structural optimization with the similar
procedure.
\subsection{Computation details for the TB-LMTO calculations}
To calculate the electronic ground state properties of  MgB$_4$, MgB$_2$C$_2$ and  LiBC
we used the TB-LMTO method of Andersen.\cite{andersen84} 
The von Barth-Hedin parameterization is used for the exchange 
correlation potential within the local density approximation.
In the present calculation, we used atomic sphere approximation.  
The calculations are semi-relativistic,
i.e. except spin-orbit coupling, all other relativistic effects are
included, taking also into account combined correction terms. 
BZ {\bf k}-point integrations 
are made using the tetrahedron method on a grid of 405 (MgB$_4$), 365 (MgB$_2$C$_2$)
 and 549 (LiBC)\,{\bf k}-points in 
the irreducible part of BZ. In order to have more insight into
the chemical bonding, we have also evaluated the
crystal orbital Hamiltonian population (COHP)\cite{dronskowski93,cohp} 
in addition to the regular band structure calculations. 
COHP is the density of states weighted by the corresponding Hamiltonial matrix
elements, a positive sign of which indicating bonding character and negative 
antibonding character.

\subsection{Computational details of FPLMTO calculations}
The full-potential LMTO calculations\cite{wills} presented in this paper
are all electron, and no shape approximation to the charge density or
potential has been used. The base geometry in this computational method
consists of muffin-tin and interstitial parts.
The basis set is comprised of augmented linear muffin-tin
orbitals.\cite{oka75} Inside the muffin-tin
spheres the basis functions, charge density and potential are
expanded in symmetry adapted spherical harmonic functions together
with a radial function. Fourier series is used in the interstitial regions. In
the present calculations the spherical-harmonic expansion of the charge
density, potential and basis functions were carried out up to
$\ell$ = 6. The tails of the basis functions outside their
parent spheres are linear combinations of Hankel or Neumann functions
depending on the sign of the  kinetic energy of the basis function in
the interstitial regions. For the core-charge density, the Dirac equation
is solved self-consistently, i.e., no frozen core approximation is used. The
calculations are based on the generalized-gradient-corrected-density-functional 
theory as proposed by Perdew {\em et al.}\cite{pw96}
SO term is included directly in the Hamiltonian matrix elements
for the part inside the muffin-tin spheres.
Moreover, the present calculations make use of a so-called
multi basis, to ensure a well converged wave function. This means that
we use different Hankel or Neuman functions each attaching to its own
radial function. This is important to obtain a reliable description of the 
higher lying unoccupied states,
especially for the optial property studies. For our
elastic properties study we have used 192\,{\bf k} points and for the
optical property studies 624 {\bf k} points in the  irreducible part of BZ.

\subsection{Calculation of optical properties}
Once the energies $\epsilon_{{\bf k} n}$ and functions $|{\bf k}n\rangle$ for the $n$ bands are obtained self
consistently, the 
interband contribution to the imaginary part of the dielectric functions
$\epsilon_{2}$($\omega$) can be calculated by summing the transitions from occupied to
unoccupied states (with fixed {\bf k} vector) over BZ, weighted
with the appropriate matrix element for the probability of the transition.
To be specific, the components of $\epsilon_{2}$($\omega$) are given by
\begin{eqnarray}
\epsilon_{2}^{ij}(\omega)  & = & \frac{Ve^2}{2\pi\hbar m^2 \omega^2}
\int d^3\! k
\sum_{ n n^{\prime} }
\bigl\langle {\bf k} n \big |  p_{i} \big | {\bf k}
n^{\prime} \bigr\rangle
\bigl\langle {\bf k} n^{\prime} \big |  p_{j} \big | {\bf
k}
n \bigr\rangle \times  \nonumber \\
\label{e2}
& & f_{{\bf k}n}\, \bigl(1 - f_{{\bf k} n^{\prime}}\bigr)
\delta\bigl( \epsilon_{{\bf k} n^{\prime}} - \epsilon_{{\bf k} n} - \hbar
\omega
\bigr) ,
\end{eqnarray}
where ($p_{x}$,$p_{y}$,$p_{z}$)={\bf p} is the momentum operator and $f_{{\bf k}n}$
is the Fermi distribution.
The evaluation of matrix elements in Eq. (\ref{e2}) is done over
the muffin-tin and interstitial regions separately. Further details about the
evaluation of matrix elements are given elsewhere.\cite{alouani}
For hexagonal structure
of MgB$_2$ the dielectric function is a tensor. By an appropriate choice
of the principal axes we can diagonalized it and restrict our considerations
to the diagonal matrix elements. We have calculated the two components 
$E\|a$ and $E\|c$ of the dielectric
constants corresponding
to the electric field parallel to the crystallographic axes
$a$ and $c$, respectively.
These calculations yield the
unbroadened functions. To reproduce the experimental conditions correctly,
it is necessary to broaden the calculated spectra. The exact form of the
broadening function is unknown, although comparison with measurements
suggests that the broadening usually increases with increasing excitation
energy. Also the instrumental
resolution smears out many fine features.
To simulate these effects the
lifetime broadening was simulated by convoluting the absorptive part of the
dielectric
function with a Lorentzian, whose full width at half maximum (FWHM) is equal to
$0.01(\hbar\omega)^2$\,eV. The experimental resolution was simulated by
broadening the final spectra with a Gaussian, where FWHM is equal to 0.02 eV.
For metals, the intraband contribution to the optical dielectric tensor influence
to the lower energy part of the spectra. This has been calculated using the
screened plasma frequency obtained from Fermi surface integration with the description
in Ref.\,\onlinecite{ravi99}.

\subsection{Calculation of Elastic properties}
The hexagonal phase of MgB$_2$ has two lattice
parameters $a$ and $c$ with Bravais lattice vectors in matrix form
\[ {\bf R} = \left( \begin{array}{ccc}
\frac{\sqrt{3}}{2} & \frac{-1}{2} & 0\\
0 & 1 & 0\\
0 & 0 & \frac{c}{a}\\
\end{array} \right). \]
The FPLMTO\cite{wills} method allows total-energy calculations to be done for arbitrary
crystal structures. We can therefore apply small strains to the equilibrium
lattice, then determine the resulting change in the total energy, and from this
information deduce the elastic constants.
The elastic constants are identified
as proportional to the second-order coefficient in a polynomial fit of the
total energy as a function of the distortion parameter $\delta$.\cite{duane}
We determine linear combinations of the
elastic constants by straining the lattice vectors ${\bf R}$ according to the relation
${\bf R' = R D}$. Here ${\bf R'}$ is a matrix containing the
components of the distorted
lattice vectors and ${\bf D}$ the symmetric distortion matrix, which contains the
strain components.
We shall consider only small lattice distortions in
order
to remain within the elastic limit of the crystal.
In the following we
shall
briefly list the relevant formulas used to obtain the elastic constants for
hexagonal crystals. The internal energy of a crystal under strain, $\delta$,
can
be Taylor expanded in powers of the strain tensor with respect to the initial internal
energy of the unstrained crystal in the following way,
\begin{eqnarray}
E(V,\delta) = E(V_{0},0) + V_{0}(\sum_{i}\tau_{i}\xi_{i}\delta_{i} + 1/2\sum_{ij}c_{ij}
\delta_{i}\xi_{i}\delta_{j}\xi_{j}) \nonumber \\
+ O(\delta^{3})
\label{eqn1}
\end{eqnarray}
The volume of the unstrained system is denoted V$_0$, E(V$_0$,0) being the
corresponding total energy. In the equation above, $\tau_{i}$ is an element in
the stress tensor.
\par
Since we have five independent elastic constants, we need five different strains
to determine them. The five distortions used in the present investigation
are described below.
The first distortion 
\[D_{1} = \left( \begin{array}{ccc}
1+\delta & 0 & 0\\
0 & 1+\delta & 0\\
0 & 0 & 1+\delta\\
\end{array} \right)
\]
gives compression or expansion to the system. This preserves the symmetry
but changes the volume.
The strain energy associated with this distortion is
\[
E(V,\delta) = E(V_{0},0) + V_{0}[(\tau_{1}+\tau_{2}+\tau_{3})\delta +
\frac{1}{2} (2 c_{11}+2c_{12}+4c_{13}+c_{33})\delta^{2}]
\]
The second distortion 
\[D_{2} = \left( \begin{array}{ccc}
(1+\delta)^{-1/3} & 0 & 0\\
0 & (1+\delta)^{-1/3} & 0\\
0 & 0 & (1+\delta)^{2/3}\\
\end{array} \right)
\]
gives volume and symmetry conserving variation of $c/a$. The energy associated with
this distortion is
\[
E(V,\delta) = E(V_{0},0) + V_{0}[(\tau_{1}+\tau_{2}+\tau_{3})\delta +
 \frac{1}{9}(
c_{11}+c_{12}-4c_{13}+2c_{33})\delta^{2}]
\]
The strain matrix 
\[D_{3} = \left( \begin{array}{ccc}
\frac{1+\delta}{(1-\delta^{2})^{1/3}} & 0 & 0\\
0 & \frac{1-\delta}{(1-\delta^{2})^{1/3}} & 0\\
0 & 0 & \frac{1}{(1-\delta^{2})^{1/3}}\\
\end{array} \right)
\]
distorts the basal plane by
elongation along $a$ and
compression along $b$ in such a way that the volume is conserved. The energy associated
with this distortion is
\[
E(V,\delta) = E(V_{0},0) + V_{0}[(\tau_{1}-\tau_{2})\delta + (c_{11}-c_{12})\delta^{2})]
\]

The elastic
constant c$_{55}$ can be determined by the distortion of the lattice using the volume
conserving triclinic distortion 

\[D_{4} = \left( \begin{array}{ccc}
\frac{1}{(1-\delta^{2})^{1/3}} & 0 & \frac{\delta}{(1-\delta^{2})^{1/3}}\\
0 & \frac{1}{(1-\delta^{2})^{1/3}} & 0\\
\frac{\delta}{(1-\delta^{2})^{1/3}} & 0 & \frac{1}{(1-\delta^{2})^{1/3}}\\
\end{array} \right)
\]
The energy change associated with
this distortion is
\[
E(V,\delta) = E(V_{0},0) + V_{0}[\tau_{5}\delta + (2c_{55})\delta^{2}]
\]
The fifth strain 
\[
 D_{5} = \left( \begin{array}{ccc}
1 & 0 & 0\\
0 & 1 & 0\\
0 & 0 & 1+\delta\\
\end{array} \right) \]
involves stretching of the $c$ axis while keeping other axes unchanged. Hence, the
hexagonal symmetry is preserved but volume is changed. The energy change associated with
this strain can be written as
\[
E(V,\delta) = E(V_{0},0) + V_{0}[\tau_{3}\delta + (\frac{c_{33}}{2})\delta^{2}]
\]
The elastic constant $c_{33}$ can be directly obtained from the above relation.
By solving the linear equations given above we have obtained all the five elastic
constants.
From pressure dependent lattice parameter measurements it is easy to obtain the
bulk modulus along the crystallographic axes. Also to quantify the mechanical anisotropy
of MgB$_2$ it is important to calculate the bulk modulus along the axes. For 
hexagonal crystals the bulk modulus along $a$ ($B_{a}$) and
$c$ ($B_{c}$) are defined as
\[
B_{a} = a\frac{dP}{da} = \frac{\Lambda}{1+\alpha}
\]
and
\[
B_{c} = c\frac{dP}{dc} = \frac{B_{a}}{\alpha}
\]
where
$\Lambda = 2(c_{11}+c_{12})+4c_{13}\alpha + c_{33}\alpha^{2}$ and
\[ \alpha = \frac{c_{11}+c_{12}-2c_{13}}{c_{33}-c_{13}}
\]
The calculated bulk modulus along the crystallographic
axes obtained from these relations are compared with the experimental results in Sec.\,\ref{sec:resdis}. 

\section{Results and discussion}
\label{sec:resdis}
\subsection{Electronic structure}
\label{electronic}
The calculated electronic band structure of MgB$_2$ is given in Fig.\,\ref{fig:bnd1}.
The interesting feature of this band structure is that a doubly degenerate, nearly flat,
bands are present just above the $E_{F}$ in the $\Gamma-$A direction in Fig.\,\ref{fig:bnd1}
and cut the $E_{F}$ along the K$-\Gamma$ direction. 
These
bands give rise to nearly cylindrical, hole-like Fermi surfaces around the
$\Gamma$ point,\cite{kortus01} indicating that the transport properties are 
dominated by the hole carriers in the plane where B atoms exist. 
These bands are incompletely filled bonding $\sigma$ bands with
predominantly boron $p_{x,y}$ character.
The $p_{z}$ bands (they are mainly in the unoccupied state and having finite
contribution along M-$\Gamma$ direction at the VB in Fig.\,\ref{fig:bnd1})
are derived from the intralayer $\pi$ bonding orbitals which also 
have interlayer couplings between adjacent atomic orbitals in the $c$ direction.
Our earlier study\cite{ravi95} on superconducting La$_3$X (X = Al,Ga, In, Tl) 
compounds show that the presence of a flat band
in the vicinity of $E_{F}$ gives large $T_{c}$ (For La$_3$In addition of carbon 
gives a stable La$_3$InC compound for which $E_{F}$ is brought to the pseudogap and
results in a non-superconducting state.)
The flat band feature is also present in the recently discovered superconducting YNi$_2$B$_2$C
compound.
As a working hypothesis, we believe that this flat-band feature plays an important role 
for the superconductivity
in MgB$_2$. It is worth to recall that the calculated\cite{kong01} electron-phonon 
interaction strength also shows a large value in the $\Gamma$-A direction where the flat-band
feature is seen. Moreover, zone boundary phonon calculations show that this band feature
is very sensitive to the $E_{2g}$ mode (B-bond stretching).
We observed that the top of this flat-band feature is around 0.54\,eV above $E_{F}$.
Our calculations show that addition of around 0.32 electron to MgB$_{2}$ will bring
$E_{F}$ to the top of this energy band (assuming rigid band filling). Thus, if 
electrons are responsible for the superconductivity one can expect
enhancement of $T_c$ on electron doping. 
Slusky {\em et al.}\cite{slusky01}
have reported the role of electron doping on the superconductivity in Mg$_{1-x}$Al$_x$B$_2$
phase for which it is found that the $T_{c}$ drops smoothly up to $x$ = 0.1 and beyond $x$ = 0.25 the 
superconductivity is completely destroyed. This indicates that holes are responsible
for the superconductivity in MgB$_2$ an observation consistent with
Hall effect measurements.\cite{hall}
The electron per atom ratio for TiB$_2$ is similar to that for MgB$_2$ and hence
it seems worthwhile to make comparison between these phases. The electronic structure
of TiB$_2$\cite{ravi01} shows that $E_{F}$ is located in a pseudogap
on the DOS curve and hence the estimated electron-phonon coupling constant
is much smaller than for superconducting materials. This observation is consistent
with the experimental findings in the sense that no superconductivity is detected for
TiB$_2$ even below 1\,K.\cite{leyarovska79}
\par
The role of Mg on the band structure of MgB$_2$ can be elucidated by completely removing
the Mg atoms from the lattice and repeating the calculations for a  
hypothetical structure with only B atoms (note: using the lattice parameters for MgB$_2$). 
The calculated electronic structure for B network alone is given in Fig.\ref{fig:bnd1} along with
that of MgB$_2$. The striking difference between the two cases is in the position of the flat
band in the $\Gamma$-A direction. 
This flat band feature for the B network alone is strongly two dimensional (viz. 
very little dispersion along 
$\Gamma$-A).
This is ca. 1.8\,eV above $E_{F}$ owing to the lower number of electrons in the B network
compared with MgB$_2$. This suggests that the B-B $\sigma$ bonds are primarily responsible for
the superconductivity in MgB$_2$.
Interestingly the bands are not deformed appreciably when we remove Mg from MgB$_2$ indicating
that the electrons from Mg atoms mainly give a shift in $E_{F}$ almost like rigid-band filling.
This view point is confirmed also from the density of states study (see Fig.,\ref{fig:pdos})
which shows that the topology of the DOS profile with and without Mg atoms in MgB$_2$ are almost 
the same. But a shift in DOS is observed when Mg is removed from MgB$_2$.
\par
BeB$_2$ could be expected to have a high $T_{c}$ owing to the
lighter Be atoms which may provide larger phonon frequencies while maintaining a
similar electronic structure to that of MgB$_2$. 
Even though BeB$_2$ is isoelectronic with MgB$_2$, the recent experimental study\cite{felner01}
did not reveal any sign of superconductivity down to 5\,K. This negative finding makes it interesting to
investigate the electronic structure of BeB$_2$ in detail. 
The lattice constants for BeB$_2$ obtained\cite{tupitsyn75} by averaging experimental data for
the actual unit cell have been inferred to be $a$ = 2.94 and $c$ = 2.87 \AA. 
There are no experimental lattice
parameters available for BeB$_2$ with the AlB$_2$-type structure and hence we
have made structural optimization using the FPLAPW method. 
Table\ref{table:bulk} shows that $a$ is reduced only by about 1.6\,\%,
whereas $c$ is by about 8\,\% when Mg is replaced by Be. The anisotropic
changes of the lattice parameters can be understood from the anisotropic
bonding situation in MgB$_2$. It looks as if the strong B ($p-p$) $\sigma$ bonds within
the planes prevent more appreciable changes in $a$ whereas the weaker $\pi$
bonds along with the relatively weak ionic bonding between Be and B bring about a larger
change in the $c$. 
Consistent with the earlier calculation\cite{satta01} our calculated value of
DOS at $E_{F}$ for BeB$_2$ is smaller than that of
MgB$_2$. 
The smaller $N(E_{F})$ value results from the broad nature of band structure of
BeB$_2$ compared with that of MgB$_2$.  Owing to the smaller
volume our calculated bulk modulus for BeB$_2$ becomes larger than that of MgB$_2$ 
(see Table\,\ref{table:bulk}). 
\par
The total DOS curve for BeB$_2$ (see Fig.\,\ref{fig:tdos}) shows almost free-electron-like 
metallic feature, 
with a $N(E_{F})$ value of  
6.309\,state Ry$^{-1}$ f.u.$^{-1}$ which in turn 
consistent with the paramagnetic behavior observed experimentally.
The DOS curves of BeB$_2$ and MgB$_2$ are indeed very similar as expected due to their isoelectronic
and postulated isostructural nature. As mentioned above the DOS features
could lead one to a higher T$_c$ for BeB$_2$ than for MgB$_2$ 
(the molecular weight of BeB$_2$ is lower than that of MgB$_2$). 
Hirish predicted\cite{hirsch01}
that the charge transfer from Be to B in BeB$_2$ is less than that from Mg to B in
MgB$_2$ and that $E_{F}$ is below the regime where superconductivity
occurs. 
A closer inspection of the
band structure (see Fig.\,\ref{fig:bnd2} indicates that the key energy band,
which we believe to be
responsible for superconductivity, is broader in BeB$_2$ than in MgB$_2$ and it is also
located well above $E_{F}$.
Hence the calculation suggests that even if one stabilizes BeB$_2$ in the AlB$_2$-type structure
one can not
expect superconductivity. Our conclusion is consistent with the experimental
observation in the sense that a recent study\cite{felner01} shows
paramagnetic behavior down to 5\,K.
\par
Kortus {\it et al.}\cite{kortus01}
suggested that Ca doping should lead to an overall increase in
DOS, and also provide an additional contribution to the electron-phonon coupling.
We have therefore calculated DOS (using the optimized structural parameters)
for CaB$_2$ which shows (Fig.\,\ref{fig:tdos}) sharp
features like those found in transition metal phases resulting from enhancement in 
volume compared with MgB$_2$. Also, the calculated DOS at $E_{F}$ is 
larger than that of MgB$_2$ indicating a possibility for superconductivity.
The larger volume compared with BeB$_2$ and MgB$_2$ along with the weak
B-B interaction make the bulk modulus for this material become smaller.
The electronic
structure of CaB$_2$ (Fig.\,\ref{fig:bnd2}) shows that the key energy band
is broader than that in MgB$_2$. Also this doubly degenerate
band is well below $E_{F}$ at the $\Gamma$ point and well above $E_{F}$ at
the A point. Hence, the calculations suggest that the
probability for superconductivity in CaB$_2$ with AlB$_2$-type
structure is low. 
\par
The total DOS for SrB$_2$ (see Fig.\,\ref{fig:tdos}) predicts a pseudogap feature at $E_{F}$ (separating
bonding from antibonding states with a negligible DOS at $E_{F}$).
The electronic structure of SrB$_2$ 
(Fig.\,\ref{fig:bnd2}) shows nearly semimetallic feature.
Earlier studies\cite{ravi96,ravi01} 
indicate that materials with $E_{F}$ located at a pseudogap in DOS will have 
relatively high stability. (The situation with all bonding orbitals 
filled and all antibonding orbitals empty implies extra contribution to
stability.) Hence the calculations predict that SrB$_2$ with 
AlB$_2$ structure may be stabilized experimentally if the above criterion
works.  However, materials with
$E_{F}$ located in the pseudogap
are not expected to become superconducting\cite{ravi95} and hence the
present finding suggests that SrB$_2$ should be non-superconducting.
Compared to BeB$_2$ and MgB$_2$, the top of the VB in CaB$_2$ and 
SrB$_2$ have large nonbonding B $p$ states (see Fig.\,\ref{fig:tdos}).
This will give negative contribution to the one-electron eigen-value
sum for stability. This may be the reason why no stable CaB$_2$ and 
SrB$_2$ compounds are stabilised experimentally in the AlB$_2$ structure.
\par
The number of valence electrons per B, C atom is same for LiBC and MgB$_2$.
So, if electron per atom ratio
is the decisive factor for the superconducting behavior of MgB$_2$ one can expect superconductivity
for LiBC. Hence, we have also performed electronic structure studies
for LiBC. The calculated DOS (Fig.\,\ref{fig:tdos}) predicts insulating behavior with a band gap of
1.81\,eV. 
LiBC is an indirect bandgap insulator where the bandgap is between
the top of VB in the $\Gamma$-K direction and the bottom of CB at the M point (Fig.\,\ref{fig:bnd3}).
The establishment of insulating behavior is consistent with the experimental
observation\cite{worle95} of a very small conductivity for LiBC. 
The flat band feature present in the K-M and $\Gamma$-A
directions of the BZ just below $E_{F}$ 
around $-$0.23 eV (see Fig.\,\ref{fig:bnd3})
suggests that LiBC may be tuned to become superconducting upon hole doping.
\par
The calculated total DOS (Fig.\,\ref{fig:tdos}) for MgB$_2$C$_2$ shows a finite number of
electrons at $E_{F}$ (which has the 
same number of valence electrons per non-metal atom as MgB$_2$)
and predicts metallic behavior.
This is further confirmed by the band structure 
(Fig.\,\ref{fig:bnd3}) which shows that several bands cross the Fermi level. The
most interesting aspect of Fig.\,\ref{fig:bnd3} is that there is flat band present in the
vicinity of $E_{F}$ (around the T$-$Y direction)
similar to that in MgB$_2$ and superconducting transition metal borocarbides
(e,g, YNi$_2$B$_2$C, LuNi$_2$B$_2$C). Consequently we suggest superconductivity with
relatively high $T_{c}$ for MgB$_2$C$_2$.
A more detailed analysis of the band structure
shows that the narrow band in the T$-$Y direction near $E_{F}$
is stemming from the C $p_{z}$ electrons. 
The calculated density of states at $E_{F}$ for MgB$_2$C$_2$ is
10.98\,states Ry$^{-1}$ f.u.$^{-1}$ which is higher than for MgB$_2$.
Except for the structural data\cite{worle94}
no other information on the physical properties of MgB$_2$C$_2$ are available 
experimentally which may be a promising candidate for high $T_{c}$ superconductivity.
\par
If it is the boron layers that are responsible for the superconductivity, one could expect
superconductivity in MgB$_4$. 
However, the calculated total
DOS for MgB$_4$ shows (Fig.\,\ref{fig:tdos}) only features which point towards insulating
behavior and hence  
superconductivity is not expected for this material.

\subsection{Superconductivity}
Let us first look for similaritiies between MgB$_2$ and other superconducting
materials. 
There appears to be different correlations\cite{asokamani89} between superconducting transition
temperature and the average electronegativity ($\eta$) for high $T_{c}$ cuprates and
conventional superconductors. Conventional
superconductors take $\eta$ value between 1.3 and 1.9 
whereas that for
high T$_c$ cuprates falls in the range 2.43 to 2.68.\cite{nishimoto00} The $\eta$ value
for MgB$_2$ is 1.733 which places this material in the category of the conventional superconductors.
However, using the proposed correlation\cite{takagi94} between
superconducting transition temperature and the value of electronic specific heat coefficient
($\gamma$) for various superconductors
we find that $\gamma$ for MgB$_2$ falls in the region of the high T$_c$ cuprates (viz. 
MgB$_2$ has a low value of $N(E_{F})$ as well as high T$_c$).
In the conventional superconductors (A15-type) with A$_3$B composition, the 
mutually orthogonally permuted linear ..A-A-A.. chains 
are believed to be responsible for the superconductivity.
In MgB$_2$, the B atoms in the zig-zag chains are believed to be playing an important 
role for superconducting behavior.  For higher $T_{c}$ of superconducting
intermetallics Butler\cite{butler81} has suggested a rule of togetherness which prescribes that (for a
given crystal structure, electron-per-atom ratio and period) the electron-phonon coupling
is enhanced when the transition metal atoms are brought closer together. It is interesting to
note that the B atoms are brought close together by the strong covalent bonding in MgB$_2$.
Earlier studies\cite{ohashi89} show that $\beta$-ZrNCl is a semiconductor with a 
band gap of $\sim$3\,eV that upon Li intercalation (the electron donor element) becomes 
superconducting
with $T_{c}$ = 13\,K.\cite{yamanaka98} This similarity is present for
MgB$_2$ also, that the electron donation from Mg to B$_2$ leads to superconductivity.
Ledbetter\cite{ledbetter94} points out that the superconducting
transition temperature in high $T_{c}$ cuprates increases with increasing Debye 
temperature ($\theta_{D}$). The high $T_{c}$ in MgB$_2$ may be associated with the large $\theta_{D}$
in this material and this criterion would place MgB$_2$ with
high T$_c$ cuprates.
\par
In conventional superconductors, T$_c$ increases with decreasing $\theta_{D}$, i.e. with
lattice softening.\cite{ledbetter94} 
The calculated $\theta_{D}$ for MgB$_2$ from the elastic constants is 1016\,K which is 
found to be much higher value than in phonon mediated superconducting materials. However, 
the calculated value
of $\theta_{D}$ is comparable with the experimental values obtained from specific 
heat measurements (Table \ref{table:theta}).
The presently derived $\theta_{D}$ is exceptionally large 
indicating that some novel mechanism is responsible for the high T$_c$ in MgB$_2$.
Concerning the relationship of the purely phononic properties to superconductivity,
owing to the large $\theta_{D}$, average phonon properties 
cannot reliably be used to estimate superconducting properties for this material. 
Using the concept that temperature-dependent electronic screening arising from narrow
bands in the vicinity of $E_{F}$ causes temperature-dependent phonon-mode
frequencies (see Fig.\,\ref{fig:bnd1}), one is lead to expect that softening of phonon modes
does occur. 
Therefore, experimental temperature dependent phonon spectra for MgB$_2$ is required to establish 
whether softening of particular phonon modes is responsible for its large T$_c$.
\par
The BCS theory and its subsequent refinements based on Eliasberg equations show that high
$T_{c}$ in phonon-mediated superconductors is favored by high
phonon frequency and a large DOS at $E_{F}$.
High $T_{c}$ materials commonly display interesting peculiarities
in the phonon-dispersion curves, often in the form of
dips (i.e. softening of phonons in well-defined regions of reciprocal space)
which are particularly evident when phonon spectrum of a high-$T_{c}$ material is compared
with that of a similar low-$T_{c}$ material (such as Nb vs. Mo, TaC vs. HfC).
Assuming that this is the case for MgB$_2$ the high T$_c$ can be explained
as follows. The superconducting transition temperature for the strongly coupled
superconductors according to the McMillan's formula\cite{mcmillan} is,

\begin{eqnarray}
T_{c} = \frac{\theta_{D}}{1.45} exp [ -\frac{1.04 (1+\lambda)}{\lambda-\mu^{*}(1+0.62\lambda)}]
\end{eqnarray}
\label{eqntc}

In Eqn.\,\ref{eqntc} large $T_{c}$ can be obtained when we have a large value
for $\theta_{D}$ (which MgB$_2$ has) and the electron-phonon coupling constant $\lambda$. 
The empirical value
of the Coulomb coupling constant $\mu^{*}$ for $sp$ metals is 0.1. The McMillan-Hopfield 
expression for $\lambda = \frac{N(E_{F})<I^{2}>}{M<\omega^{2}>}$,
which enters in the exponent in the expression for T$_{c}$. 
Here, $N(E_{F})$ is smaller for MgB$_2$
than for conventional high $T_{c}$ materials, $<I^{2}>$ is the
averaged square of the electron-phonon matrix element, $<\omega^{2}>$ is the 
averaged square of the phonon frequency and $M$ is the mass of the ion involved. 
If the superconductivity occurs by phonon mediation, high $T_{c}$ in MgB$_2$ can 
be explained as follows. Selected
phonon modes may be strongly coupled to the electronic system and influence the
magnitude of $T_{c}$ to a greater extent than average phonon modes. 
In consisitent
with the above view point the lattice dynamical calculations\cite{kong01,yildrim01,liu01,an01}
reveal that the in-plane boron phonons near the zone-center are highly anharmanic
with significant non-linear contribution to the electron-phonon coupling.
If such a situation
occurs, the softening of certain phonons in combination with the light mass of the 
constituents may lower the phonon contribution ($M<\omega_{ph}^{2}>$) and in turn enhance
the electron-phonon coupling constant and hence $T_{c}$. This may explain why
MgB$_2$ possesses high $T_{c}$ despite the lower value of $N(E_{F})$. 
\par
Now we will compare MgB$_2$ with relevant superconducting
material.
Among the other transition metal diborides, superconductivity with $T_{c}$=9.5\,K
has been observed in TaB$_2$ recently.\cite{kaczorowski01} Our electronic structure
studies\cite{ravi01} show that the $E_{F}$ is located in a peak in the DOS profile in TaB$_2$.
This feature along with the large $N(E_{F})$ value of 12.92\,states/(Ry f.u.) will
explain the large $T_{c}$ in this material.
As there is substantial B $p$ states present at $E_{F}$ in MgB$_2$ similar to superconducting
RNi$_2$B$_2$C, it may be worthwhile to compare these two cases. An experimental soft
x-ray emission spectroscopy study\cite{shin95} on superconducting YNi$_2$B$_2$C
and non-superconducting LaNi$_2$B$_2$C along with bandstructure calculation\cite{mattheiss94}
indicate that the superconductivity appears only when the broad B $p$ bands are located
at $E_{F}$.
The main difference between the electrons involved in transport properties of
RNi$_2$B$_2$C and MgB$_2$ is that the former has a remarkable DOS peak at $E_{F}$
dominated by Ni 3$d$ states with almost equal proportions of all five Ni 3$d$ states
and also involves some rare-earth $d$ and B, C $sp$ admixture whereas MgB$_2$ does not show 
any peak feature at $E_{F}$.
From the accurate analysis of measured\cite{walti01} specific heat over a wide temperature
range the estimated $\gamma$ for MgB$_2$ is 
estimated as 5.5\,mJ/mol K$^{2}$.
From our calculated $N(E_{F})$ value we have derived the electronic contribution to
the specific-heat coefficient without electron-phonon mass enhancement yield
a value of 1.73\,mJ/mol K$^{2}$. From this value along with the experimental $\gamma_{exp}$
we have estimated the value of electron-phonon coupling constant using the relation
$\gamma_{exp}$ = $\gamma_{th}(1+\lambda)$ which gave $\lambda$ = 2.17. The large $\lambda$ value
indicates MgB$_2$ is a strongly coupled superconductor. This strong coupling
as well as the low mass of B can explain the high $T_{c}$ in MgB$_2$.
\par
Using the calculated plasma frequency ($\Omega_{px}$) in the $ab$ plane (7.13\,eV) along with
$N(E_{F})$ = 0.71\,states/(eV f.u.) 
the in-plane Fermi velocity $\nu_{Fx}$ has been obtained as 
\[
\nu_{Fx}=\sqrt
{\frac
{\Omega_{px}^{2}}
{4\pi e^{2} N(E_{F})}} = 8.9\times10^{7}\,cm/s.
\]
Using the superconducting
gap $\Delta$=3.53$k_{B}T_{c}$/2 = 6\,meV we have calculated the coherence
length ($\xi$) and the field penetration depth ($\Lambda$) 
\[
\xi = \frac{\hbar \nu_{Fx} }{\pi\Delta} = 307\,\AA; \Lambda = \frac{c}{\Omega_{px}} = 276\,\AA.
\]
The experimental upper critical field, $H_{c2}(T)$, thermodynamic critical field,
$H_{c}(T)$, and critical current, $J_{c}$, indicate that MgB$_2$ is a type-II
superconductor.\cite{finnemore01}
\par
Now we will try to understand the effect of pressure on the superconductivity in MgB$_2$.
Addition of Mg to the boron sublattice gives 15\% shortening
of the $c$ axis compared to that in graphite.\cite{baskaran01} This chemical pressure brings
the B semimetal to superconducting state. 
As the transport properties of MgB$_2$ is from the holes, it is suggested that the
superconductivity may be understood within the formalism developed for high-$T_{c}$
cuprate superconductors.\cite{hirsch01} This predicts a positive pressure
coefficient for $T_{c}$ as a result of the decreasing intraplane B-B distance with
increasing pressure. 
However, the $T_{c}$ is experimentally\cite{tomita01} found to decrease with pressure
at a rate of $-$1.11\,K/GPa.
The RVB theory,\cite{baskaran01} on the other hand, predicts a decrease in $T_{c}$ with 
increasing chemical pressure, 
consistent with the experimental findings.\cite{tomita01,lorenz01,saito01}
Linear response calculations\cite{kong01} show that the B-bond stretching modes have unusually 
strong coupling to electrons close
to $E_{F}$ at the top of the bonding quasi-two-dimensional B $\sigma$ bonds.
Therefore, when the compressibility is larger within the $ab$ plane than along $c$, one can expect 
a large variation in the superconductivity with pressure (because the flat bands
with $p_{x,y}$ characters broaden faster by compression). 
Our calculated elastic property for MgB$_2$ shows large anisotropy with easy compression along $c$. 
Hence, the key energy band (which is sensitive to the B bond stretching) will broaden
slowly with pressure and $T_{c}$ will slowly decrease with increasing pressure, consistent with
experimental results.
\par
From the FPLAPW calculations we have estimated the electric field gradient
($V_{zz}$) at the Mg and B site as $-$0.249 $\times$10$^{21}$ 
and 2.047 $\times$ 10$^{21}$V/m$^{2}$, respectively.
Using the calculated $V_{zz}$ along with the nuclear quadrapole moment for
$^{11}$B (0.037 b)\cite{shirley75} we can calculate the NMR quadrapole coupling frequency
($\nu_{q}$) by means of the relation
\[
\nu_{q} = \frac{3eQV_{zz}}{2hI(2I-1)}
\]
using I = 3/2 as the nuclear spin quantum number for $^{11}$B. This gave 
$\nu_{q}$ = 915 kHz that is in 
good agreement with 828 kHZ obtained by Gerashenko 
{\em et al.}\cite{gerashenko01} and 835 $\pm$ 5 kHz obtained by 
Jung {\em et al.} from first order quadrapole perturbed NMR spectrum.\cite{jung01}

\subsection{Chemical bonding in MgB$_2$}
Similar to the C-C distances in the graphite structure, the distance between the boron 
planes in MgB$_2$ is about twice
the intraplanar B-B distance and hence the B-B bonding is strongly anisotropic.
A more quantitative assessment of the bonding situation  
in MgB$_2$ can be obtained from the partial DOS (Fig.\ref{fig:pdos})  
which demonstrates that the B $s$ states are
hybridized with the B $p$ state in VB. This shows strongly bonded
$sp^{2}$ hybrids in the $ab$ plane. The Mg $s$ electrons 
contribute very little to the VB and are mainly reflected in the unoccupied state. 
Hence Mg donates electrons to the boron layers. VB carries predominantly B 2$p$
character formed from two distinct sets [$\sigma$ (p$_{x,y}$) and $\pi$ ($p_{z}$)]
of bonds. The B 2$s$ electrons are well localized and their contribution at $E_{F}$ is minor.
From the orbital projected DOS (Fig.\,\ref{fig:pdos}) it is clear that 
B $p_{x}$ and $p_{y}$ characters are
mainly dominating at $E_{F}$. This suggests that the $p-p$ $\sigma$ bonding between
the boron atoms has a significant influence on superconductivity. 
It is worth to recall
that the recently found\cite{yamanaka98} superconductor Na$_{0.29}$HfNCl 
(T$_c$ $\approx$ 25\,K) has N $p_{x}$ and $p_{y}$ characters at $E_{F}$.\cite{weht99}
B $p_{z}$ states are present in a wide energy range and dominating at the
bottom of CB.
\par
The simplest way to
investigate the bonding situation between two interacting atoms in solid is to inspect 
the complete COHP between them, taking all valence orbitals into  account. 
In the upper panel of Fig\,\ref{fig:pdos} shows COHP for the B-B and Mg-B bonds.
An interesting aspect of this illustration is that VB is filled up with
bonding orbitals (negative value of COHP) and the antibonding orbitals are some 
$\sim$3\,eV above $E_{F}$.
Bonding state electrons from both B-B and Mg-B bonds are found at $E_{F}$.
The B $s-s$ bonding states are found mainly at the bottom of VB around
$-$8.5\,eV. The B $p-p$ $\sigma$ bonding states dominate at
the top of the VB region around $-$2\,eV. Note that the COHP
values for the Mg-B bonds are much smaller than for the B-B bonds indicating that the
B-B bond is much stronger than the Mg-B bonds. In order to quantify the bonding interactions
in MgB$_2$ we have integrated the COHP curve and obtained the values $-$4.36, $-$1.23 and $-$0.307\,eV 
for B-B, Mg-B and Mg-Mg bonds respectively. This further confirms that the
B-B bonds are the strongest in MgB$_2$ which in turn consistent with the
derived elastic properties (see Sec.\,\ref{subsec:elas}).
\par
In order to further illustrate the bonding situation in MgB$_2$ a charge density
plot for the (110) plane is shown in Fig.\,\ref{fig:charge}. This illustration shows
a low electron accumulation between Mg and B as well as a very
low electron population at the Mg site (much lower than for a neutral Mg atom).
These findings are a clear indication of ionic bonding between Mg and B. 
The large electron accumulation between the B atoms and their strongly
aspherical character indicate strong covalent
interaction between the B atoms as also found by our examination of partial DOS and COHP.  
The more or less homogeneous charge distribution 
between the Mg atoms suggest an appreciable degree of metallic bonding between them. 
i.e. apart from strong ionic and covalent bonding in MgB$_2$
the band structure shows features similar to the $sp$ metals.
Hence, MgB$_2$ is a typical example of a mixed bonded solid.

\subsection{Elastic properties}
\label{subsec:elas}
Structural parameters for BeB$_2$, CaB$_2$, SrB$_2$ are not available
experimentally and we have therefore made structural optimization for these diborides along
with the MgB$_2$ by total energy minimization. The calculated total energy
variation as a function of $c/a$ and volume for MgB$_2$ is shown in Fig.\,\ref{fig:vol}a
and \ref{fig:vol}b respectively.
The optimized structural parameters along with the bulk
modulus, its pressure derivative and DOS at $E_{F}$ for
all these diborides are given in Table\,\ref{table:bulk}. The calculated equilibrium
volume for MgB$_2$ is found to be in excellent agreement with the experimental
value,\cite{jones54} and the corresponding $c/a$ value is only 0.42$\%$ smaller 
than the experimental value (see Fig.\,\ref{fig:vol}b and Table\,\ref{table:bulk}).
These results indicate that density functional theory works well for this material. 
The calculated $N(E_{F}$ value for MgB$_2$ is larger than that for other diborides
considered in the present study and this may be one of the reasons for the 
superconductivity in MgB$_2$. 
\par
By fitting the total-energy vs. volume curve in Fig.\,\ref{fig:vol} to the 
universal equation of state we have
obtained the bulk modulus ($B_{0}$) and its pressure derivative ($B_{0}^{\prime}$)
as 1.50\,Mbar and 3.47, respectively. The derived $B_{0}$
is found to be in excellent agreement with the value 1.51\,Mbar obtained
from high resolution x-ray powder diffraction measurements\cite{vogt01} 
and also in good agreement with other results listed in Table\,\ref{table:bulk}
Furthermore $B_{0}$ = 1.5\,Mbar obtained in this way is in good agreement with the value 1.509\,Mbar
estimated from the calculated single-crystal elastic constants by means of the relation
\[
B= \frac{2(c_{11}+c_{12})+4c_{13}+c_{33}}{9}.
\]
The electron per atom ratio of 
TiB$_2$ is same as that of MgB$_2$ and hence this material should also be considered.
In TiB$_2$ $E_{F}$ is located in a pseudogap 
and hence $N(E_{F}$ is small resulting in non-observation of superconductivity.
The calculated bulk modulus for TiB$_2$ 
is larger than that for diborides considered in the present study.
This is an effect of the filling of the bonding states in the bands and this aspect
has been discussed in Ref.\,\onlinecite{ravi01}.
\par
From various distortions we have calculated all the five single-crystal elastic
constants for MgB$_2$ (see Table\,\ref{table:elastic}). Unfortunately there are no experimental 
elastic constants available since a suitable 
single crystal of MgB$_2$ has so far not been obtained.
We have therefore made comparison with the data for TiB$_2$ available experimentally\cite{spoor97}. 
Despite the layered crystal structure of MgB$_2$ the high pressure total energy studies
of Loa and Syassen\cite{loa01} suggested isotropic compressibility
and they concluded that the intra- and the interlayer bonding are of similar strength. 
Another high pressure study (upto 8\,Gpa) by Vogt {\em et al.}\cite{vogt01} concluded that
there are small anisotropies in the mechanical properties. 
The isothermal compressibility measurements of MgB$_2$ by synchrotron x-ray
diffraction revealed\cite{prassides01} a stiff tightly-packed incompressible nature
with only moderate anisotropy between intra- and interlayer bonds.
The neutron diffraction measurements by Jorgensen {\em et al.}\cite{jorgensen01} at high pressures
concluded with highly anisotropic mechanical properties for MgB$_2$.
As the experimental results are mutually inconsistent, theoretical studies
of the compressibility may be helpful in resolving the ambiguities.
\par
From the calculated single-crystal elastic constants we have derived the bulk moduli
along the crystallographic directions using the relations given in Sec.\,\ref{sec:details}
and the results are listed in Table\,\ref{table:elastic}. 
The compressibility study\cite{prassides01} by synchrotron radiation show 
isothermal interlayer compressibility, dln$c$/d$P$ at zero pressure is
1.4 times the inplane compressibility, dln$a$/d$P$. 
However the high pressure neutron diffraction studies\cite{jorgensen01} show
that dln$c$/d$P$=1.64 dln$a$/d$P$ which is much closer to our calculated relationship
dln$c$/d$P$=1.79 dln$a$/d$P$ (see Table\,\ref{table:elastic}. The most recent
high pressure measurement\cite{goncharov01} by hydrostatic pressure upto 15\,GPa shows a
large anisotropy with the relationship dln$c$/d$P$=1.875 dln$a$/d$P$.
The higher compressibility along $c$ than along $a$ can be understood as follows. 
There is strong B $p_{x,y}-p_{x,y}$ covalent hybridization
along $a$ in MgB$_2$ and hence the bulk modulus along $a$ ($b$) is large. 
There is significant ionic contribution to the bonding between Mg and B along $c$. Usually
an ionic bond is weaker than a covalent bond and hence the bulk modulus is smaller 
along $c$ than along $a$. 
The large anisotropy in the compressibility is also consistent
with the fact\cite{jorgensen01} that the 
thermal expansion of $c$ is about twice as that of $a$.
Moreover, substitution\cite{slusky01} of Al for Mg site decreases $c$ at a rate
approximately twice that of $a$ indicating the anisotropic nature of bonding.
\par
It is possible to visualize the anisotropy in the elastic properties from the 
curvature of the total energy with respect to
length changes in an arbitrary direction. From the elastic compliance constants
($s_{ij}$) it is possible to derive the
directional bulk modulus K, using the following relation:\cite{nye85}
\begin{eqnarray}
\frac{1}{K} = (s_{11}+s_{12}+s_{13})-(s_{11}+s_{12}-s_{13}-s_{33})l_{3}^{2}
\end{eqnarray}
where $l_{3}$ is the direction cosine. Thus obtained
directional dependent bulk modulus is shown in Fig.\,\ref{fig:bg}a. A useful
surface construction is one that shows the directional dependence of
Young's modulus (E), which for hexagonal symmetry can be defined as 
\begin{eqnarray}
\frac{1}{E} = (1-l_{3}^{2})^{2}s_{11}+l_{3}^{4}s_{33}+l_{3}^{2}(1-l_{3}^{2})
(2s_{13}+s_{44})
\end{eqnarray}

Thus derived directional-dependent Young's modulus (Fig.\,\ref{fig:bg}b) also shows
large anisotropy, $E$ along $a$ being about 65$\%$ larger than along $c$. The anisotropic
nature of bonding behavior reflected in the elastic properties is consistent
with the charge density analysis. The marked anisotropic compressibility of MgB$_2$ will 
lead to different pressure effects on different phonon modes and is also
more likely to lead to pressure induced changes in the electronic structure at $E_{F}$.
This information is valuable in testing the predictions of competing
models\cite{hirsch01,baskaran01} for the mechanism of superconductivity.
\par
The anisotropy in the plastic properties of materials can be
studied from the directional dependent shear stress and the amount of shear.\cite{boas50}
So, it is interesting to study the directional dependence of
shear in MgB$_2$. We have calculated the directional dependent
shear modulus (G) from the elastic constants using the
relation
\begin{eqnarray}
\frac{1}{G} = s_{44} + (s_{11}-s_{12}-\frac{s_{44}}{2}) (1-l_{3}^{2}) + 2(s_{11}+s_{33}
-2s_{13}-s_{44})(1-l_{3}^{2})l_{3}^{2}
\end{eqnarray}
The directional dependent shear modulus (see Fig.\,\ref{fig:bg}c) also shows
large anisotropy and the G along the $a$ axis is around 42$\%$ higher
than that along $c$. It should be noted that large shear value is present
in between the basal plane and perpendicular to basal plane. From the calculated
directional dependent bulk modulus and the shear modulus
one can calculate the directional dependence of sound velocity
and hence the characteristic temperature of the material. The calculated
characteristic temperature given in Fig.\,\ref{fig:bg}d shows
anisotropic nature due to the anisotropy in the bonding behavior.
The calculated average shear modulus ($G_{av}$), elastic-wave velocities ($\nu$) and $\theta_{D}$
obtained from our single crystal elastic constants are listed in
Table\,\ref{table:theta}. The calculated $\theta_{D}$ for MgB$_2$ is much higher 
than the experimental values and the higher value of $\theta_{D}$ indicates that 
higher phonon frequencies play an important role for the high $T_{c}$ of this
material.
The existence of phonon modes at very high energies as 97\,meV in
MgB$_2$ is experimentally identified by neutron-inelastic-scattering measurements.\cite{sato01,osborn01}
One of the reasons for the discrepancy between the $\theta_{D}$ obtained from
the elastic constants and the specific heat measurements is as follows.
For example, the experimental electrical resistivity and specific heat measurements
on TiSi$_2$ gave $\theta_{D}$ = 560-664\,K, whereas 
elastic constants measurements gave $\theta_{D}$ = 700\,K (that is in
close agreement with $\theta_{D}$ = 722\,K obtained from our theoretical elastic
constant calculations\cite{ravi98}). So, one can expect good agreement when comparing
our $\theta_{D}$ with that obtained from elastic constants measurement.

\subsection{Optical properties}
\label{sec:optic}
Optical properties studies are of fundamental importance, since
these involve not only the occupied and unoccupied parts of the electronic structure
but also carries information on the character of the bands. In order to elucidate the anisotropy in the
optical properties of MgB$_2$ the calculated imaginary parts of dielectric tensor
for $E\|a$ and $E\|c$ are shown in Fig.\,\ref{fig:optic}. The important feature
conveyed by the interband transition shown in Fig.\ref{fig:optic} is that there is negligible
contribution to $\epsilon_{2}(\omega)$ in $E\|c$ below 3.8\,eV. The interesting
aspect for MgB$_2$ is that even though the interband contribution to the optical
spectra is highly anisotropic the calculated intraband contribution show nearly
isotropic behavior. This originates from the close values of the calculated 
plasma frequencies; 7.13 and 6.72\,eV for
in-plane and perpendicular to the plane, respectively. These values are in 
excellent agreement with $\omega_{p,x}$= 7.02\,eV and $\omega_{p,z}$= 6.68\,eV 
obtained by full-potential LMTO calculation.\cite{kong01}.
The intraband contribution to the optical dielectric tensor have been calculated
similar to our earlier study.\cite{ravi99}
The calculated $\epsilon_{2}(\omega)$ spectra which include both inter- and
intraband contributions are shown in the upper panel of fig.\ref{fig:optic}.
From these $\epsilon_{2}(\omega)$ spectra one can derive all linear optical properties.
Unfortunately, there are no experimental spectra available for MgB$_2$.

\section{SUMMARY}
\label{sec:con}
The present study shows a common origin between the superconductivity in 
rare earth transition metal borocarbides and MgB$_2$
that there is a flat band present in the vicinity of Fermi level 
and also the B atoms are primarily involved in electron-phonon coupling in these materials. 
Owing to the lack of single crystals the anisotropy in physical properties of
MgB$_2$ are not studied experimentally. We predicted large anisotropies
in the optical and mechanical properties.
From detailed electronic structure studies we have arrived at the following
conclusions.\\
\noindent
1. The bonding behavior in MgB$_2$ has been explained by analyses of site, angular
momentum and orbital projected density of states as well as charge density and crystal overlap
Hamiltonian population. These analyses establish a 
mixed bonding behavior with ionic bonding between Mg and B, covalent
bonding between B atoms and metallic bonding between Mg and Mg like that in $sp$
metals. \\

\noindent
2. We identified a large anisotropy in the mechanical properties of MgB$_2$ 
from our calculated elastic constants, consistent with the anisotropy in the
bonding behavior and high pressure neutron diffraction measurements. Consequently, pressure
can influence the bands in different directions of the BZ in an unusually different manner 
and hence the physical properties.

\noindent
3. Two degenerate flat bands have been identified near $E_{F}$ in the $\Gamma$-A 
direction of BZ. These degenerate B $p_{x,y}$ bands are considered as the 
key in the realization of the high temperature superconductivity of MgB$_2$.\\

\noindent
4. The role of Mg in MgB$_2$ is to donate electrons to B atoms and hence to shift $E_F$ such
that it lies very closer to the flat band, the feature which we believe important for superconductivity.\\

\noindent
5. We found similarity in the electronic structures of MgB$_2$ 
LiBC and MgB$_2$C$_2$. Therefore superconductivity is expected in MgB$_2$C$_2$
and hole doped LiBC. On the other hand the electronic structure of 
BeB$_2$, CaB$_2$, SrB$_2$ and MgB$_4$ suggest a low probability for superconductivity.\\

\noindent
6. Because of the nearly flat bands in the vicinity of $E_{F}$ (which will introduce temperature
depending electronic screening of the phonon-mode frequencies),
we expect temperature-dependent phonon-mode softening or large anisotropy in the phonon modes.\\

\noindent
7. The calculated boron NMR frequency is found to be in very good agreement with the experimental
studies.\\

\noindent
8. As our calculated Debye temperature for MgB$_2$ is much larger than for other
superconducting intermetallics, we believe that selected phonon modes can be 
strongly coupled to the
electronic system and thus influence the magnitude of $T_{c}$ to a greater extent than the
average phonon correlations may indicate.\\

\noindent
9. If the rigid band-filling approximation works and also the
electrons are responsible for superconductivity in MgB$_2$, our calculation suggests that a 
doping of 0.32 electrons will enhance $T_{c}$. However, the experimental observation of
reduction in $T_{c}$ by electron doping in MgB$_2$ 
indicate that the holes are responsible for superconductivity.\\

\acknowledgements
PR is grateful for the financial support from the Research
Council of Norway. 
Part of these calculations were carried out on the
Norwegian supercomputer facilities (Programme for Supercomputing). PR wishes to acknowledge  
Prof.O.K.Andersen, Prof.Ove Jepsen, Dr.Florent Boucher, Dr.John Wills, Prof.K.Schwarz 
and Prof.Peter Blaha for providing some of the programmes used in this study,
and Dr.Anna Delin and Dr.Lars Fast for useful communications.

\begin{figure}
\caption{
Crystal structures of MgB$_2$, MgB$_2$C$_2$, LiBC and MgB$_4$. Legends to the
different kinds of atoms are given on the illustration.
}
\label{fig:str}
\end{figure}
\begin{figure}
\caption{
Band structure of MgB$_2$ and hypothetical B$_2$ in the AlB$_2$-type framework.
For the B$_2$ substructure we have used the lattice parameters
of MgB$_2$.
}
\label{fig:bnd1}
\end{figure}
\begin{figure}
\caption{
Lower panel: Total and site projected DOS for MgB$_2$ and hypothetical B$_2$
obtained from the FPLMTO method. Middle panel: Angular momentum
and orbital projected DOS for MgB$_2$ obtained from the FPLAPW method 
Upper panel: Crystal overlap Hamiltonian population (COHP) between Mg-Mg, Mg-B and
B-B obtained from the TBLMTO method.
}
\label{fig:pdos}
\end{figure}
\begin{figure}
\caption{
Calculated total DOS for  BeB$_2$, CaB$_2$, SrB$_2$ obtained from the
FPLAPW method and that for MgB$_4$, LiBC
and MgB$_2$C$_2$ obtained from the TBLMTO method. For BeB$_2$, CaB$_2$,
and SrB$_2$ we have assumed the AlB$_2$-type structure and the
optimized structural parameters.
}
\label{fig:tdos}
\end{figure}
\begin{figure}
\caption{
The band structure of BeB$_2$, CaB$_2$ and SrB$_2$ with  AlB$_2$-type
structure as obtained using the optimized lattice parameters.
}
\label{fig:bnd2}
\end{figure}
\begin{figure}
\caption{
The band structure of LiBC and MgB$_2$C$_2$ as obtained with the TBLMTO
method using experimental lattice parameters.
}
\label{fig:bnd3}
\end{figure}
\begin{figure}
\caption{
Valence charge density in the (110) plane for MgB$_2$. 
Corner atoms are Mg and the others are B.
Both plots have 35 contours between 0 and 0.25 electrons/a.u.$^3$.
}
\label{fig:charge}
\end{figure}
\begin{figure}
\caption{
The variation of the total energy with unit-cell volume (lower panel) and  c/a 
(upper panel) for
MgB$_2$ (AlB$_2$-type structure) as obtained from the FPLMTO calculations.
}
\label{fig:vol}
\end{figure}
\begin{figure}
\caption{
Calculated directional-dependent (a) bulk, (b) Young's, (c) shear moduli and
the characteristic temperature (d) 
for MgB$_2$ as obtained from the calculated single crystal elastic constants.
}
\label{fig:bg}
\end{figure}
\begin{figure}
\caption{
The calculated optical dielectric tensor for MgB$_2$ obtained from with and
without intraband contribution obtained from the FPLMTO calculation.
}
\label{fig:optic}
\end{figure}

\begin{table}
\caption{Calculated  lattice parameters ($a$ and $c$  are in \AA), $c/a$ ratio, bulk modulus 
(B$_{0}$ in Mbar), its pressure derivative
(B$_{0}^{\prime}$) and  density of state at the Fermi level [$N(E_{F}$) in states Ry$^{-1}$
f.u.$^{-1}$]  for AlB$_2$-type compounds} 
\begin{tabular}{lcccccc } 
Compound & $a$ & $c$ & $c/a$ & $B_{0}$    & $B_{0}^{\prime}$ & $N(E_{F}$)   \\
\hline
BeB$_2$ & 2.886  & 3.088  & 1.027  & 1.93  & 3.47  &  6.309     \\
MgB$_2$	& 3.080  & 3.532  & 1.147  & 1.50  & 3.50  &  9.98   \\
 ...    & ...    & ...    & ...    & 1.20$\pm$0.05\tablenotemark[1] & ...& ...    \\
 ...    & ...    & ...    & ...    & 1.39\tablenotemark[2] & ...&  ...   \\ 
 ...    & ...    & ...    & ...    & 1.51\tablenotemark[3] & ...&  ...   \\ 
 ...    & ...    & ...    & ...    & 1.47\tablenotemark[4] & ...&  ...   \\ 
 ...    & ...    & ...    & ...    & 1.40$\pm$0.06\tablenotemark[5] & ...&  ...   \\ 
CaB$_2$	& 3.397  & 4.019  & 1.183  & 1.34  & 3.42  &  10.837   \\
SrB$_2$	& 3.456  & 4.193  & 1.213  & 1.05  & 2.28  &  4.732 \\
TiB$_2$	& 3.070  & 3.262  & 1.060  & 2.13  & 2.10  &  4.27  \\
\end{tabular}
\tablenotetext[1] {From synchrotron XRD by Prassides {\em et al.} \protect\cite{prassides01}.} .
\tablenotetext[2] { High pressure XRD measurement by Vogt {\em et al.}\protect\cite{vogt01}.}
\tablenotetext[3] { Calculated from FPLAPW method by Vogt {\em et al.}\protect\cite{vogt01}.} 
\tablenotetext[4] { Calculated from pseudopotential method by Bohnen {\em et al.}\protect\cite{bohnen01}.} 
\tablenotetext[5] { Calculated from FPLAPW method by Lao and Syassen\protect\cite{loa01}.} 
\label{table:bulk}
\end{table}
\begin{table}
\caption{Average shear modulus ($G_{av}$ in Mbar), longitudinal and transverse elastic wave velocity
($\nu_l$ , $\nu_k$ in m/s) and Debye temperature ($\Theta_{D}$ in K) obtained from single crystal 
elastic constant($\Theta_{D}$ in K) } 
\begin{tabular}{lccccc } 
Compound & $G_{av}$ & $\nu _l $ & $\nu _m$  &  $\Theta_{D}$(calc.) & $\Theta_{D}$(exp.)\\    
\hline
MgB$_2$	          & 1.146 & 10612  & 7276 & 1016  & 750$\pm30$\tablenotemark[1] \\
...               & ...   & ...    & ...  & ...   & 746\tablenotemark[2]  \\ 
...               & ...   & ...    & ...  & ...   & 800\tablenotemark[3]  \\
...               & ...   & ...    & ...  & ...   & 920\tablenotemark[4]  \\
TiB$_2$(exp.)$\tablenotemark[5]$ & 2.407 & 7061   & 5105 & 743   & ...       \\
\end{tabular}
\tablenotetext[1] { From specific heat measurements by Bud'ko {\em et al.} \protect\cite{budko01}.}
\tablenotetext[2] { From specific heat measurements by W\"{a}lti {\em et al.} \protect\cite{walti01}.}
\tablenotetext[3] { From specific heat measurements by Kremer {\em et al.} \protect\cite{kremer01}.}
\tablenotetext[4] { From specific heat measurements by Wang {\em et al.} \protect\cite{wang01}.}
\tablenotetext[5] {Calculated from the single crystal elastic constants of 
Spoor {\em et al.}\protect\cite{spoor97}.}
\label{table:theta}
\end{table}
\begin{table}
\caption{ The single crystal elastic constants ($c_{ij}$ in Mbar) and bulk modulus
values along $a$  and $c$ ($B_a$ and $B_c$ in Mbar)  } 
\begin{tabular}{lccccccc  } 
Compound & $c_{11}$ &  $c_{12}$ &  $c_{13}$ &   $c_{33}$ & $c_{44}$ &    $B_a$ & $B_c$\\    
\hline
MgB$_2$	& 4.380  & 0.430  &0.329   & 2.640  & 0.802  &   5.406 & 3.006     \\
  ...   & ...    & ...    & ...    & ...    & ...    &   4.1$\pm0.2$\tablenotemark[1] & 2.92$\pm0.12$ \\
  ...   & ...    & ...    & ...    & ...    & ...    &   6.25\tablenotemark[2] & 3.33 \\
TiB$_2$(exp.)\tablenotemark[3] & 6.6  & 0.48  & 0.93  &  4.32   & 2.6 & 8.512 & 5.527    \\
\end{tabular}
\tablenotetext[1] {From synchrotron XRD by Prassides {\em et al.} \protect\cite{prassides01}.}
\tablenotetext[2] {From hydrostatic high pressure synchrotron XRD by Goncharov {\em et al.} \protect\cite{goncharov01}.}
\tablenotetext[3] {Calculated from the single crystal elastic constants of Spoor {\em et al.}\protect\cite{spoor97}.}
\label{table:elastic}
\end{table}

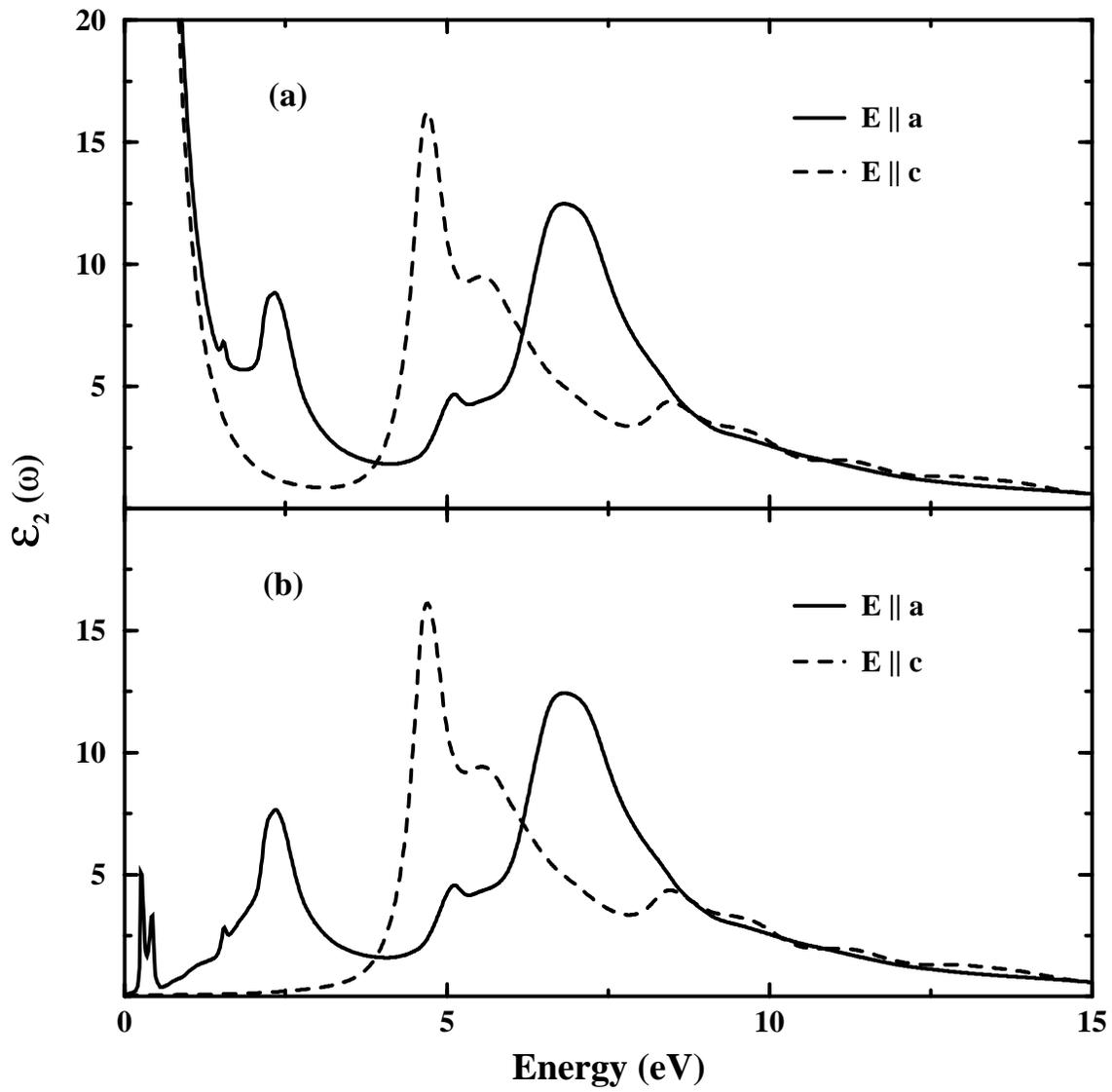
\end{document}